\documentclass[%
 aip,
 apl,%
 amsmath,amssymb,
preprint,%
]{revtex4-1}

\usepackage{graphicx}
\usepackage{dcolumn}
\usepackage{bm}
\usepackage{epstopdf}
\usepackage{natbib}
\usepackage{color}
\usepackage{subfigure}
\usepackage{setspace}

\renewcommand{\Re}{\textit{Re}}
\newcommand{\Rs}{\textit{Re}_{s}}

\newcommand{\Amp}{A}

\def\ddp#1_#2{\frac{\partial #1}{\partial #2}}
\def\ddpp#1_#2{\frac{\partial^2 #1}{\partial #2^2}}
\def\ddppz#1_#2_#3{\frac{\partial^2 #1}{\partial #2 \partial #3}}
\def\ddppp#1_#2{\frac{\partial^3 #1}{\partial #2^3}}
\def\d#1_#2{\frac{\displaystyle \mbox{d}#1}{\displaystyle\mbox{d}#2}}
\def\dd#1_#2{\frac{\mbox{d}^2#1}{\mbox{d}#2^2}}
\def\ddd#1_#2{\frac{\mbox{d}^3#1}{\mbox{d}#2^3}}

\def\bbf#1{\mathchoice{\hbox{\boldmath$\displaystyle#1$}}
{\hbox{\boldmath$\textstyle#1$}} {\hbox{\boldmath$\scriptstyle#1$}} {\hbox{\boldmath$\scriptscriptstyle#1$}} }

\newcommand{\xpart}{\bbf{X}_{p}}

\begin{document}


\title{Transport of inertial particles by viscous streaming in arrays of oscillating probes}
\author{Kwitae Chong}
\affiliation{Mechanical and Aerospace Engineering, University of California, Los Angeles, Los Angeles, CA 90095\\}
\author{Scott D. Kelly}
\affiliation{Mechanical Engineering and Engineering Science, The University of North Carolina at Charlotte, Charlotte, NC 28223\\}
\author{Stuart T. Smith}
\affiliation{Mechanical Engineering and Engineering Science, The University of North Carolina at Charlotte, Charlotte, NC 28223\\}
\author{Jeff D. Eldredge}
\email[Author to whom correspondence should be addressed.  Electronic mail: ]{eldredge@seas.ucla.edu} \affiliation{Mechanical and Aerospace
Engineering, University of California, Los Angeles, Los Angeles, CA 90095\\}
\date{\today}

\begin{abstract}
A novel mechanism for the transport of microscale particles in viscous fluids is demonstrated. The mechanism exploits the trapping of such particles
by rotational streaming cells established in the vicinity of an oscillating cylinder, recently analyzed in previous work. The present work explores a
strategy of transporting particles between the trapping points established by multiple cylinders undergoing oscillations in sequential intervals. It
is demonstrated that, by controlling the sequence of oscillation intervals, an inertial particle is effectively and predictably transported between
the stable trapping points. Arrays of cylinders in various arrangements are investigated, revealing a quite general technique for constructing
arbitrary particle trajectories. The timescales for transport are also discussed.
\end{abstract}

\maketitle

\section{Introduction}\label{sec:intro}
\begin{figure}[t]
\centering
   \subfigure[]{
    \includegraphics[scale=0.50]{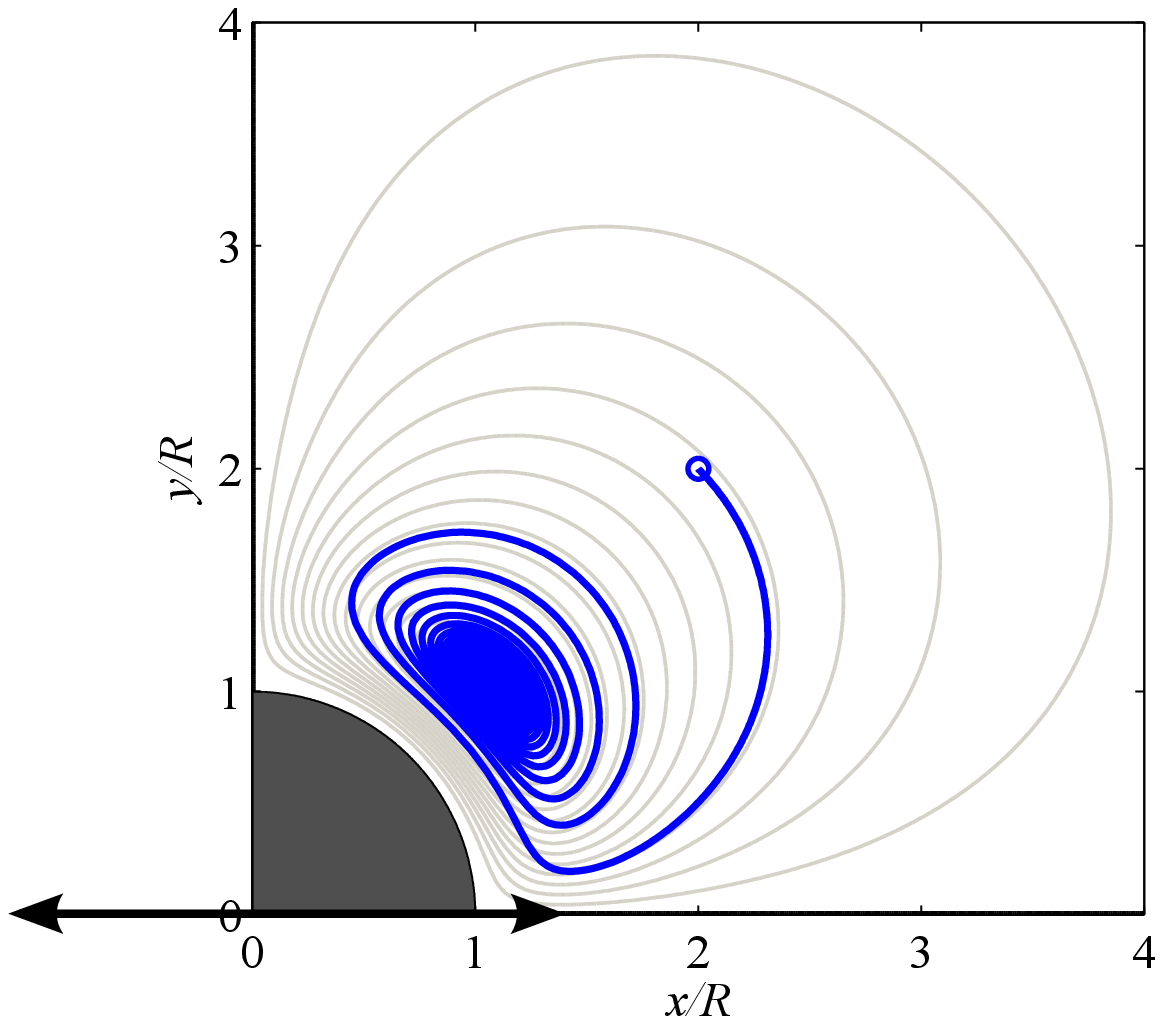}
     }
    \subfigure[]{
    \includegraphics[scale=0.32]{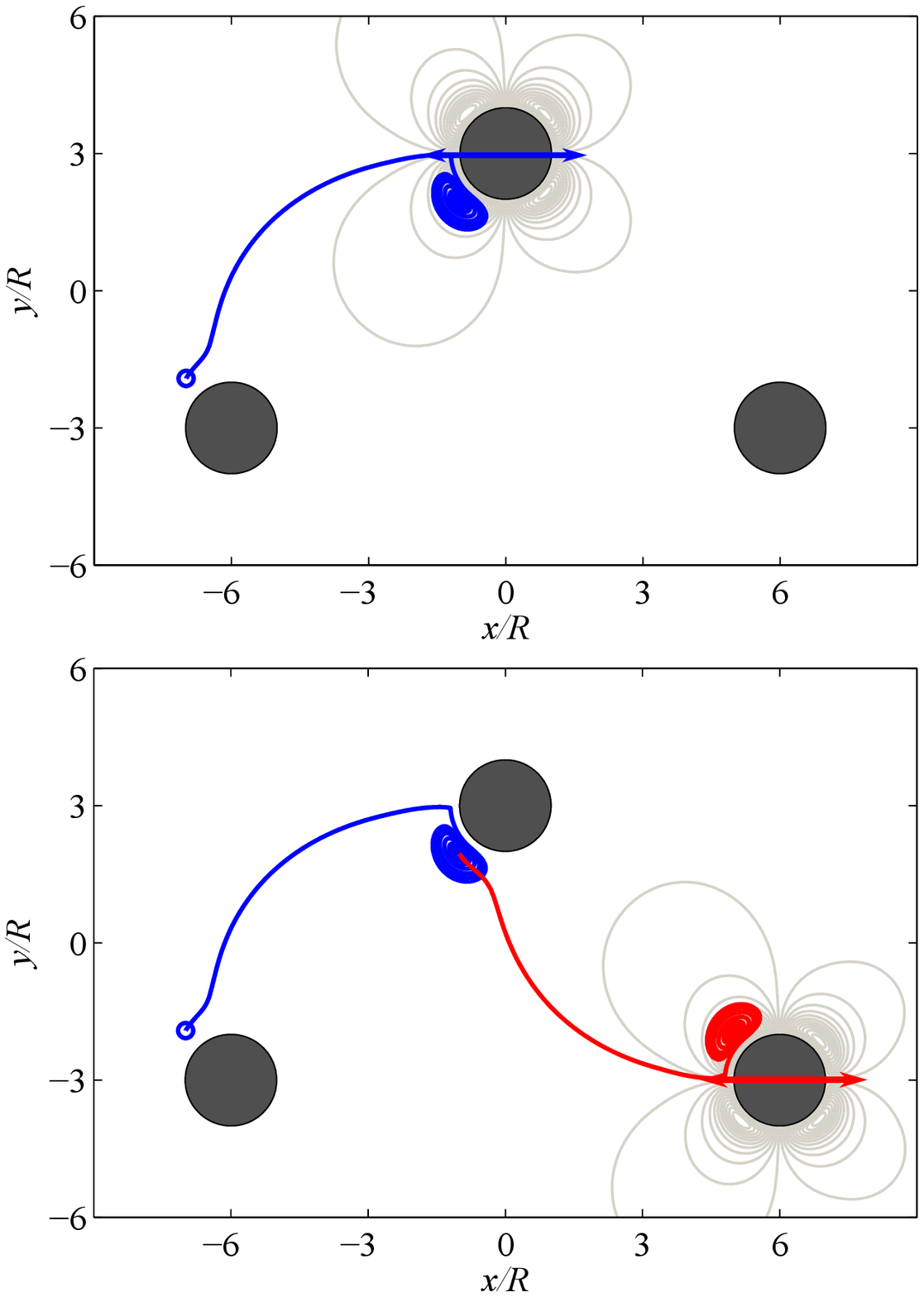}
     }
   \caption{ (a) Inertial particle trajectory (blue line) inside streaming cell based on time-averaged Lagrangian streamline (gray line).
 Initial location is depicted with circle. Reprinted with permission from Chong \emph{et al.}, Phys. Fluids 25, 033602 (2013). Copyright 2013, American Institute of Physics. (b) Schematic of transport of inertial particle in arrays of cylinders. Inertial particle is transported
   from the lower left cylinder to the upper cylinder (top), and from the upper cylinder to the lower right cylinder (bottom) by sequential
   oscillation of cylinders.
    }  \label{fig:intro}
\end{figure}

Technological advances in cellular engineering, medical diagnostics and microfluidics have motivated the need to separate, focus and transport
micromete-size particles. In recent years, a variety of mechanisms have been exploited for such purposes, including dielectrophoresis
\citep{weitz,Renaud}, optofluidics \citep{kim,dholakia,chiou2005}, inertial hydrodynamics \citep{diCarlo3,diCarlo4,schroeder}, among others. Another
mechanism, viscous streaming, has recently been demonstrated as a viable candidate for particle manipulation
\citep{hilgenfeldt:2j,hilgenfeldt:4j,hilgenfeldt:5j,Lutz:1j,jiexu1,jiexu2,chong_streaming1}. Viscous streaming  is a steady secondary flow that
develops from the coherent non-linear interactions within an oscillatory flow. For example, a cylinder vibrating rectilinearly normal to its axis
will generate four steady rotational cells in its vicinity. Previous studies have demonstrated experimentally that such cells can trap inertial
particles within a domain of influence. In recent work \citep{chong_streaming1}, we numerically investigated the mechanisms by which this trapping
occurs (see figure \ref{fig:intro}(a)). For the first time, we revealed that an inertial particle will be trapped regardless of its relative size,
density and Reynolds number, provided that the particle is smaller than the characteristic size of the oscillator. As particle size decreases, the
trapping time increases. The final trapping location is insensitive to particle size and density.

In principle, this trapping mechanism can form the basis for a transport strategy, wherein multiple cylinders vibrate in sequence to move particles
from one trapping point to another, as shown in figure \ref{fig:intro}(b). However, it is difficult to demonstrate this experimentally. In the
present work, we extend our numerical investigation to various arrays of oscillating cylindrical probes to explore particle transport between them.

\section{Methodology}\label{sec:methodology}
Consider an inertial particle whose radius and density are $a$ and $\rho_p$, respectively, in a streaming flow generated by one or more cylinders
with equal radius $R$. Each cylinder is prescribed with a sinusoidal oscillation along the $x$-axis, with displacement relative to the mean position
described by $\Amp \sin \Omega t$ (as shown for a single cylinder in figure \ref{fig:intro}(a)) with amplitude $\Amp$ and angular frequency $\Omega$
in an otherwise quiescent fluid with viscosity $\mu$ and density $\rho_{f}$. Each cylinder oscillates for a finite interval in time; the oscillation
intervals of each cylinder are sequential and non-overlapping. The Reynolds number for this flow is defined as $\Rs = \rho_{f}\Omega R^2/\mu$ and
oscillation amplitude is assumed to be small, $\epsilon = \Amp/R \ll 1$. In our previous work, we showed that it was sufficient to rely on a one-way
coupling, wherein the flowfield obtained analytically in the absence of particle was used to obtain forces on the particle; the particle's effect on
this flowfield is ignored. Here, we use the same strategy. However, analytical solutions for the flow in an array of cylinders are generally not
possible, so instead, we rely on high-fidelity numerical computations with the Viscous Vortex Particle Method (VVPM) \citep{jdevvpm:1j}. This method
has been validated on many flows, including the streaming due to a single cylinder.

In this work, it is important to consider the wide range of length and time scales. For example, vorticity is largely confined to a thin Stokes layer
$\delta_{AC} \sim Re^{-1/2}R$ and inertial particles are transported inside a streaming cell of size $\delta_{DC} \sim O(R)$. It takes $O(\Re)$
oscillations for vorticity to diffuse over this streaming cell, $R^2/\nu=Re T/{2\pi}$ (where $T=2\pi/\Omega$ is the period of oscillation) and takes
many more oscillations for inertial particles to travel around this cell with streaming velocity $V_s=\epsilon A \Omega$, i.e., the transport
timescale is of order $R/\epsilon A \Omega=\epsilon^{-2}T/{2\pi}$ for small $\epsilon$. The role of underlying oscillatory flow in determining the
force on the particle cannot be ignored. Both our previous study \citep{chong_streaming1} and the present one focus on streaming Reynolds number
($\Rs = \rho_{f}V_s R/\mu = \epsilon^2\Re$) less than unity, for which the streaming topology is relatively simple. Due to the importance of the wide
range of time and length scales, it is crucial that all scales are sufficiently resolved in the simulation.

\begin{figure}[t]
\centering
   \subfigure[]{
    \includegraphics[width=0.4\textwidth]{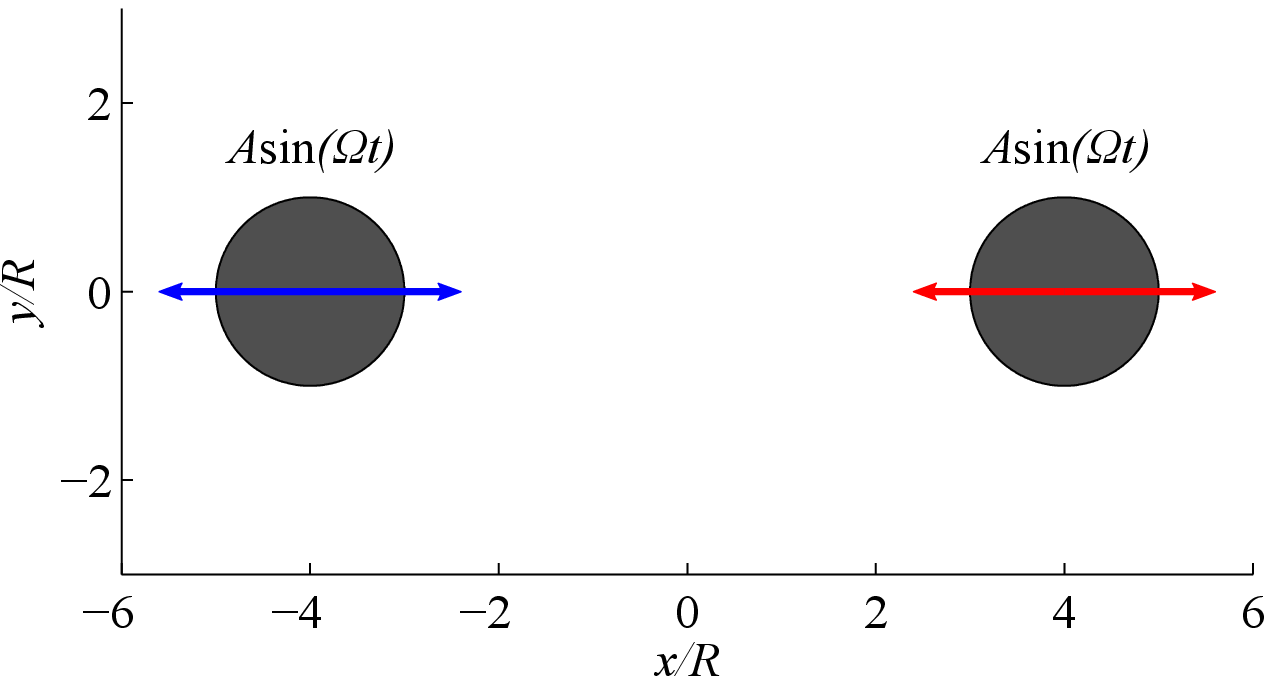}
     }
  \subfigure[]{
    \includegraphics[width=0.4\textwidth]{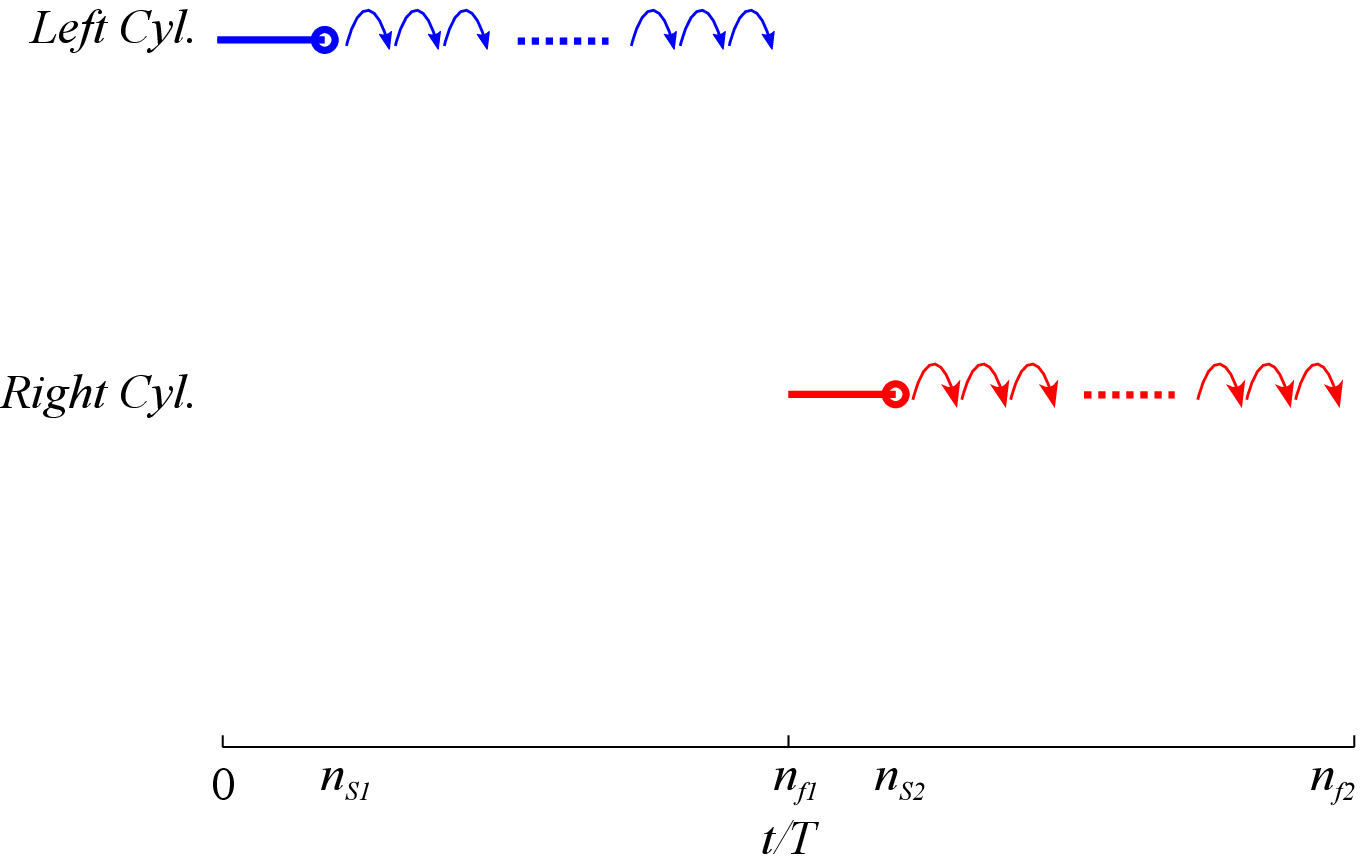}
   }\\
   \caption{ Schematic of two oscillating cylinders and the sequence of their oscillations. (a) Oscillating two-dimensional cylinders with amplitude $\Amp$ and
   frequency $\Omega$. (b) Sequence of oscillating cylinders.}  \label{fig:twoCyl}
\end{figure}

\subsection{Particle transport strategy}\label{sec:strategy}
In this work, we consider transport in various cylinder arrays. The basic strategy can be described  with two cylinders, as shown in figure
\ref{fig:twoCyl}. The left cylinder, located at $(-4R,0)$, is initially oscillated. After an inertial particle is trapped inside the streaming cell
of the left cylinder, the oscillation of this cylinder is stopped and the oscillation of the right cylinder located at $(4R,0)$ is started
immediately to instigate the trapped inertial particle to move to the newly established streaming cell of the right cylinder. Figure
\ref{fig:twoCyl}(b) illustrates the oscillation sequences. The flowfield is generated by oscillating the left cylinder until $t/T=n_{s1}$ when the
streaming flow field has reached the steady state (blue solid line). Then, the flowfield from one oscillating cycle after $t/T=n_{s1}$ is repeatedly
recycled to numerically integrate the inertial particle trajectory until it is finally trapped at $t/T=n_{f1}$ (blue dotted line). The right cylinder
starts to oscillate at $t/T=n_{f1}$ and the flowfield is computed until $t/T=n_{s2}$ when the streaming flow reaches steady state again (red solid
line). Then, one oscillating cycle after $t/T=n_{s2}$ is again recycled. Meanwhile, the trapped inertial particle is allowed to move freely until it
is finally trapped inside the center of the newly established streaming cell at $t/T=n_{f2}$ (red dotted line). In this manner, the high-fidelity
solution of the background flow is only required as long as necessary to account for the transient influence from each cylinder's initial motion.

In this study, attention is restricted to an inertial particle of radius $a/R=0.175$ and neutral density $\rho_{p}/\rho_{f}=1$, in a flow of Reynolds
number $Re=40$ and amplitude $\epsilon = 0.2$. The motion of an inertial particle inside this flow is calculated from the Maxey-Riley equation (MR
equation) \citep{riley83} with an additional Saffman lift force \citep{saffman}. The validity of utilizing the MR equation in such a flow was
established in our previous work \citep{chong_streaming1}. In that work, it was shown that the Basset history force is negligible in this regime.
Here, that force is ignored for simplicity. It was also shown in \citet{chong_streaming1} that the buoyant force and Fax\'en correction terms cause
the inertial particle velocity to deviate from the fluid particle velocity. Once the particle deviates from the fluid particle trajectory, the
Saffman lift acts upon it in the presence of a shear flow, and provides the main mechanism for particle trapping. In contrast, the Stokes drag
resists this deviation by driving the relative velocity between inertial particle and fluid to zero.

Compared to the equation (14) of our previous work \citep{chong_streaming1}, the particle transport equation used here also includes wall effect
forces, such as lubrication force and an elastic collision force from direct contact between particle and wall, since the inertial particle
 approaches the cylinder surface on certain trajectories. Both wall effect treatments are adapted from the work of by \citet{hunt2012}. In particular,
 the Stokes drag term is modified with a factor $\lambda$, in which lubrication lubrication force and Stokes force are blended by $\lambda =
0.5{\delta}^{-1}H\left(\delta/\delta_{sl}\right) + 1-H\left(\delta/\delta_{sl}\right)$, where $H\left(z\right) = \left[1+\exp(10(z-1))\right]^{-1}$
represents a smoothed and shifted Heaviside function, and $\delta = h/(2a)$ denotes the gap distance, $h$, between wall (surface of cylinder) and
particle surface, normalized by particle diameter; $\delta_{sl}$ is chosen as 0.5. As the gap distance decreases, lubrication force --- manifested by
the first in $\lambda$ --- dominates, whereas the term recovers the Stokes drag force as the gap distance increases.

We also include an elastic collision force, $F\left(\delta/\delta_{ss}\right)e_{d}W_{o}$, between inertial particle and cylinder, where
$F\left(z\right) = \left( \exp(-z) - \exp(-1) \right)/(1- \exp(-1)) $ for $0 \leq z \leq 1$ and $F\left(z\right) = 0$ for $z > 1$. This force is
based on the Hertz elastic force \citep{timoshenko}. The elastic force, $W_{o} = \frac{4}{3}a^{2}E^{\ast}\left( 5\pi\rho_{p}{V_{I}}^{2}/4E^{\ast}
\right)^{3/5}$, thus becomes active only when the gap $\delta$ is less than $\delta_{ss}$ and becomes fully active when $\delta$ approaches to zero.
The critical gap $\delta_{ss}$ and the dry coefficient of restitution $e_{d}$ are set as 0.017 and 0.97, respectively \citep{hunt2012}; $V_{I}$ is
the particle impact velocity and elastic coefficient associated with each materials(particle and wall) is set as $E^{\ast} = 3 \times 10^9$ Pa.

\section{Results}\label{sec:result}
\begin{figure}[t]
\begin{center}
\includegraphics[width=0.4\textwidth]{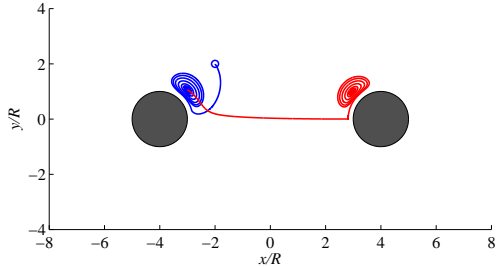}
\caption{Inertial particle trajectory between two cylinders. Initial location depicted with blue circle.}\label{fig:iner_traj_two}
\end{center}
\end{figure}

In the first scenario, depicted in figure \ref{fig:iner_traj_two}, the transport is considered between two cylinders. The inertial particle starts at
$(-2R,2R)$ and, over the interval $t/T \in [0,6784]$ (blue line), becomes trapped in the upper right streaming cell generated by the left cylinder.
Once the right cylinder starts to oscillate, the trapped inertial particle starts to move toward it, following a streamline near the line of symmetry
of the right cylinder's upper left streaming cell. The speed of the particle is relatively low since the streaming motion weakens as the inverse cube
of distance from the oscillator \citep{holtsmark54}. When it is close to the right cylinder, both the lubrication force and the collision force
become briefly active as the particle migrates upward. Finally, the particle is drawn to the center of this new streaming cell and trapped.

\begin{figure}[t]
\centering
   \subfigure[]{
    \includegraphics[scale=0.4]{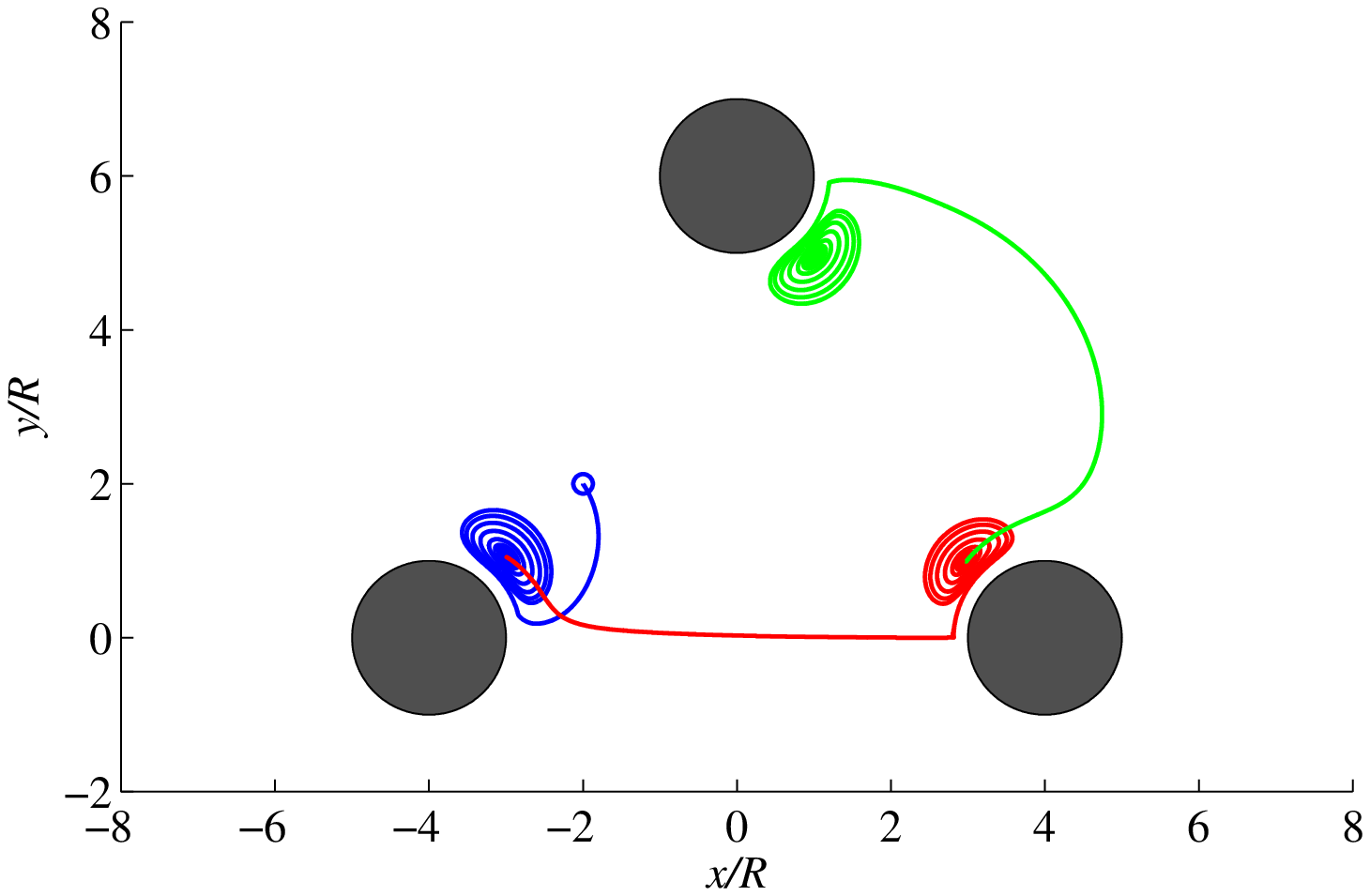}
     }
  \subfigure[]{
    \includegraphics[scale=0.4]{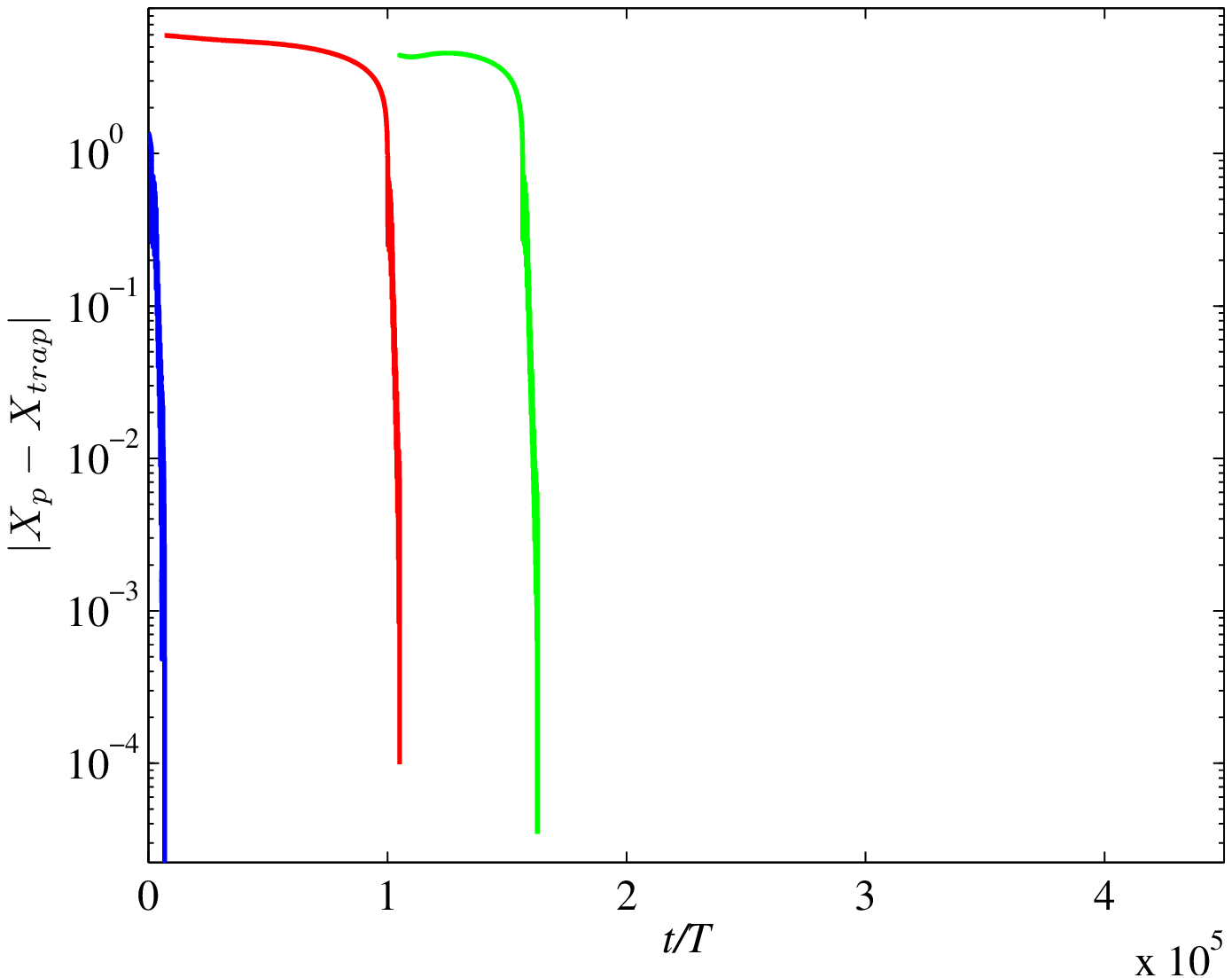}
   }\\
   \caption{(a) Inertial particle trajectory in a triangular arrangement. (b) Distance between inertial particle ($\xpart$) and trapping points
   ($\bbf{X}_{trap}$) of each respective oscillator.}  \label{fig:tri}
\end{figure}

One can also transport an inertial particle with oscillating cylinders in a triangular arrangement, as in figure \ref{fig:tri}(a). Between the lower
left and lower right cylinder, the inertial particle is transported in a similar manner as between two oscillating cylinders in figure
\ref{fig:iner_traj_two}. The trapped particle in the vicinity of the lower right cylinder starts to migrate toward and is finally trapped by the
upper cylinder, following a streamline generated by the motion of that upper cylinder during $t/T \in [105110,162754]$ (green line). Although it is
not shown in this figure, the particle trapped by the upper cylinder is transported back inside the center of the upper right streaming cell of the
lower left cylinder during $t/T \in [162755,239306]$.

\begin{figure}[t]
\centering
   \subfigure[]{
    \includegraphics[scale=0.4]{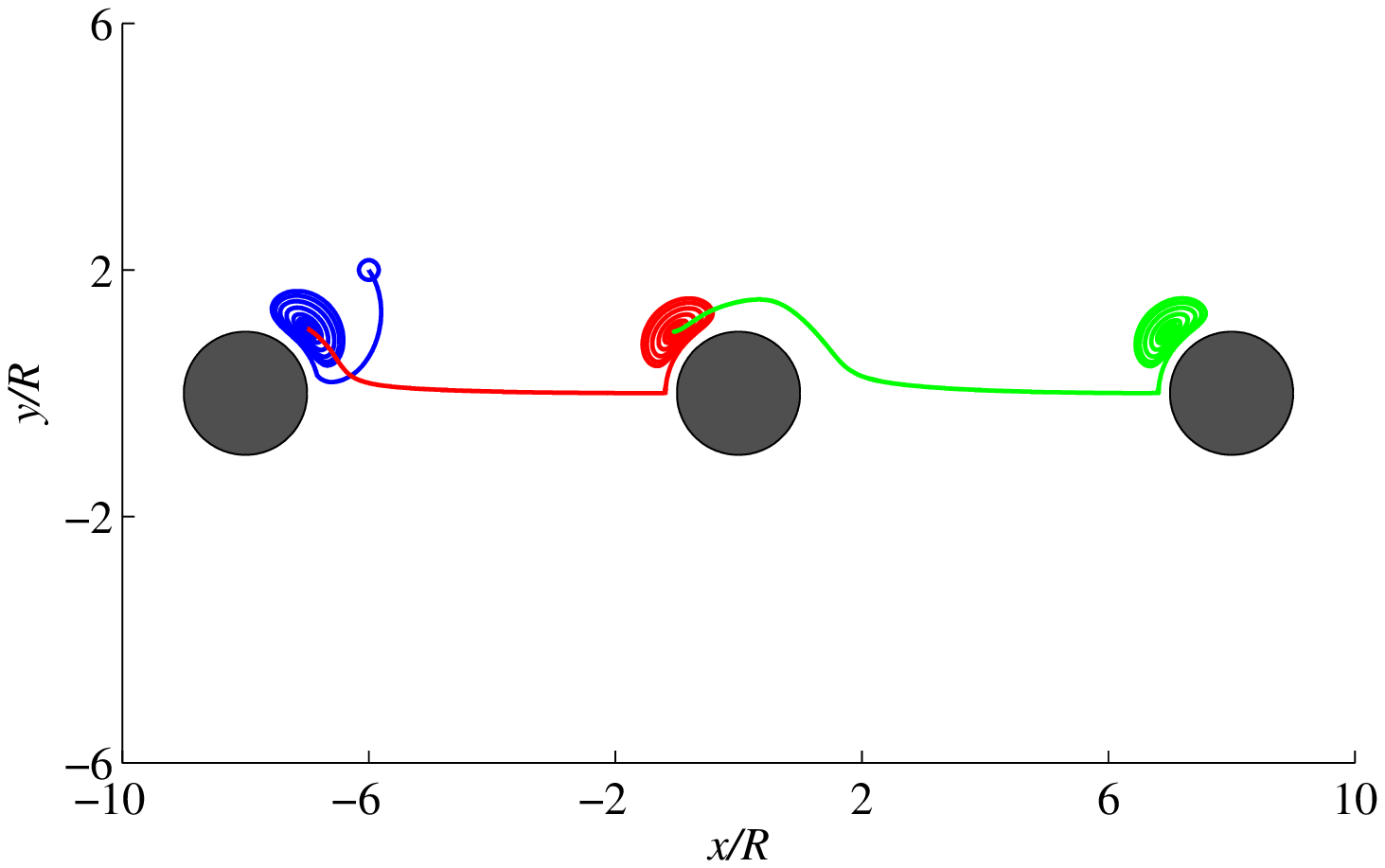}
     }
  \subfigure[]{
    \includegraphics[scale=0.4]{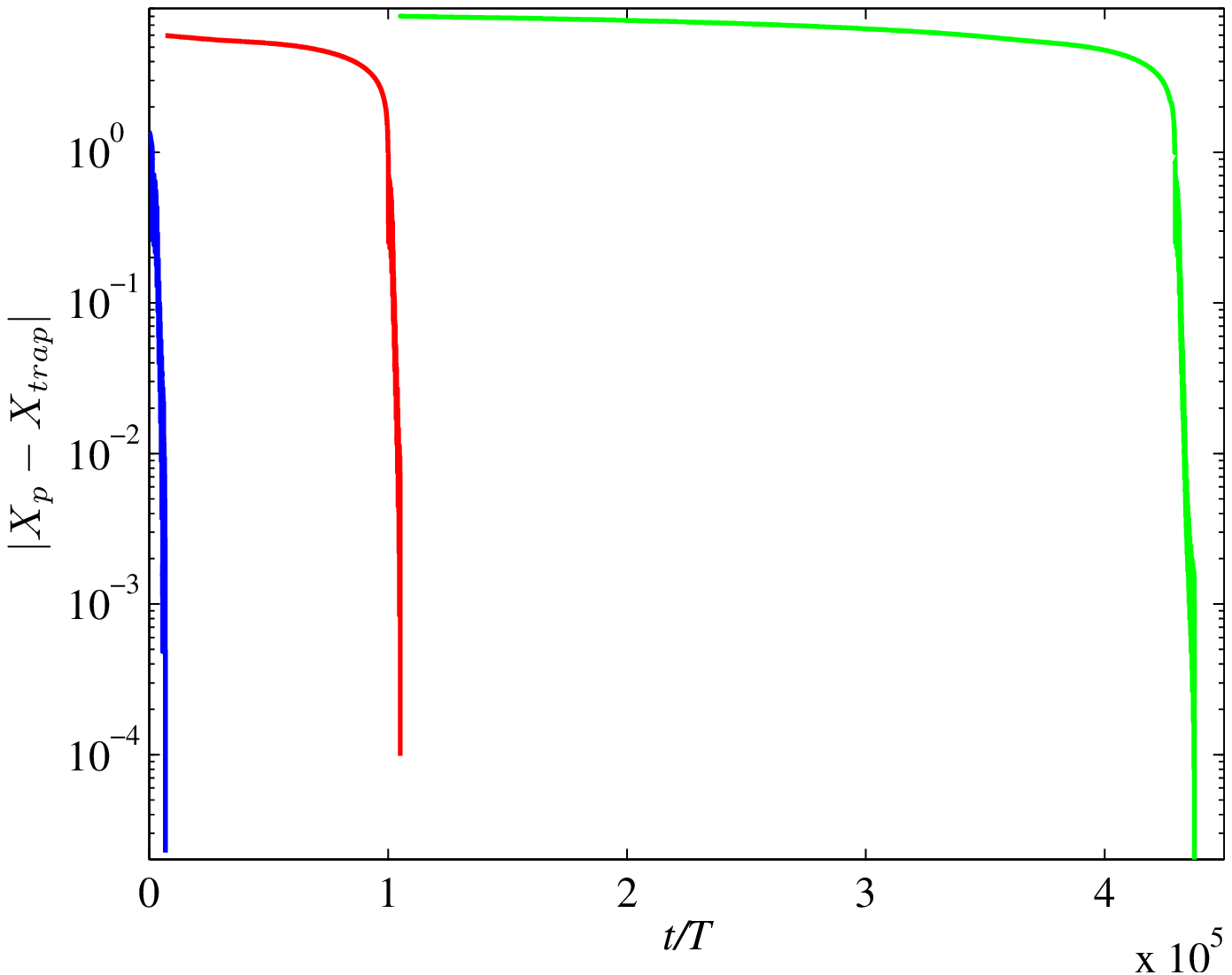}
   }\\
   \caption{(a) Inertial particle trajectory in a linear arrangement. (b) Distance between inertial particle ($\xpart$) and trapping points
   ($\bbf{X}_{trap}$) of each respective oscillator.}\label{fig:line}
\end{figure}

An inertial particle can also be transported in a linear arrangement, as shown in figure \ref{fig:line}(a). The cylinders are oscillated in sequence
from left to right. As a result, the inertial particle is initially trapped during the interval $t/T \in [0,6784]$ by the leftmost cylinder (blue
line), then migrates toward and is trapped by the center cylinder in $t/T \in [6785,105109]$ (red line), and is finally trapped by the rightmost
cylinder during the interval $t/T \in [105110,441381]$ (green line). Since the streaming flow generated by the oscillation of the rightmost cylinder
is felt weakly in the region of upper left quadrant of the center cylinder, the migration speed of the inertial particle toward the rightmost
cylinder begins relatively slowly. However, it is important to note that the streaming can induce motion even behind obstacles. It takes more than
four hundred thousand oscillations to transport the inertial particle approximately from $(-8R,0)$ to $(8R,0)$ in a reasonably straight line.

Figure \ref{fig:tri}(b) and \ref{fig:line}(b) indicate the distance between the inertial particle ($\xpart$) and trapping point ($\bbf{X}_{trap}$) of
each respective oscillator, in the configuration corresponding to panel (a) in each figure. Each particle transport interval is composed of slow
migration (plateau part) and fast trapping (plunging part). In the triangular arrangement, the second slow migration (red plateau) is slower than the
third (green plateau) despite the shorter traveling distance, since the particle on the red line of figure \ref{fig:tri}(a) follows a streamline near
the line of symmetry, but the particle on the green line follows an inner streamline with correspondingly stronger streaming velocity. In the linear
arrangement, the first two transport intervals exhibit similar behavior but the third indicates the slower migration due to the weak streaming in the
vicinity of the center cylinder.

\begin{figure}[t]
\centering
   \subfigure[]{
    \includegraphics[scale=0.4]{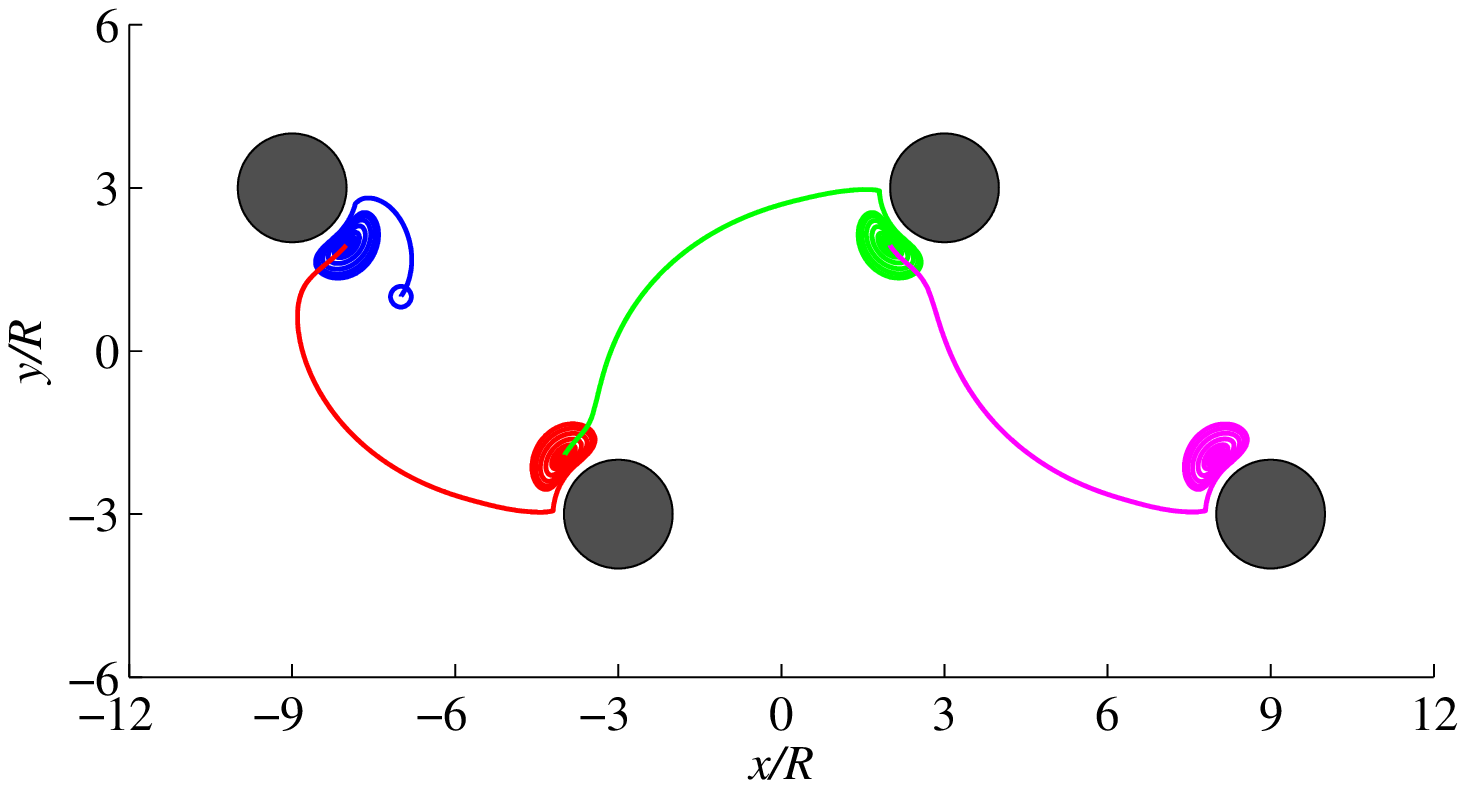}
     }
  \subfigure[]{
    \includegraphics[scale=0.4]{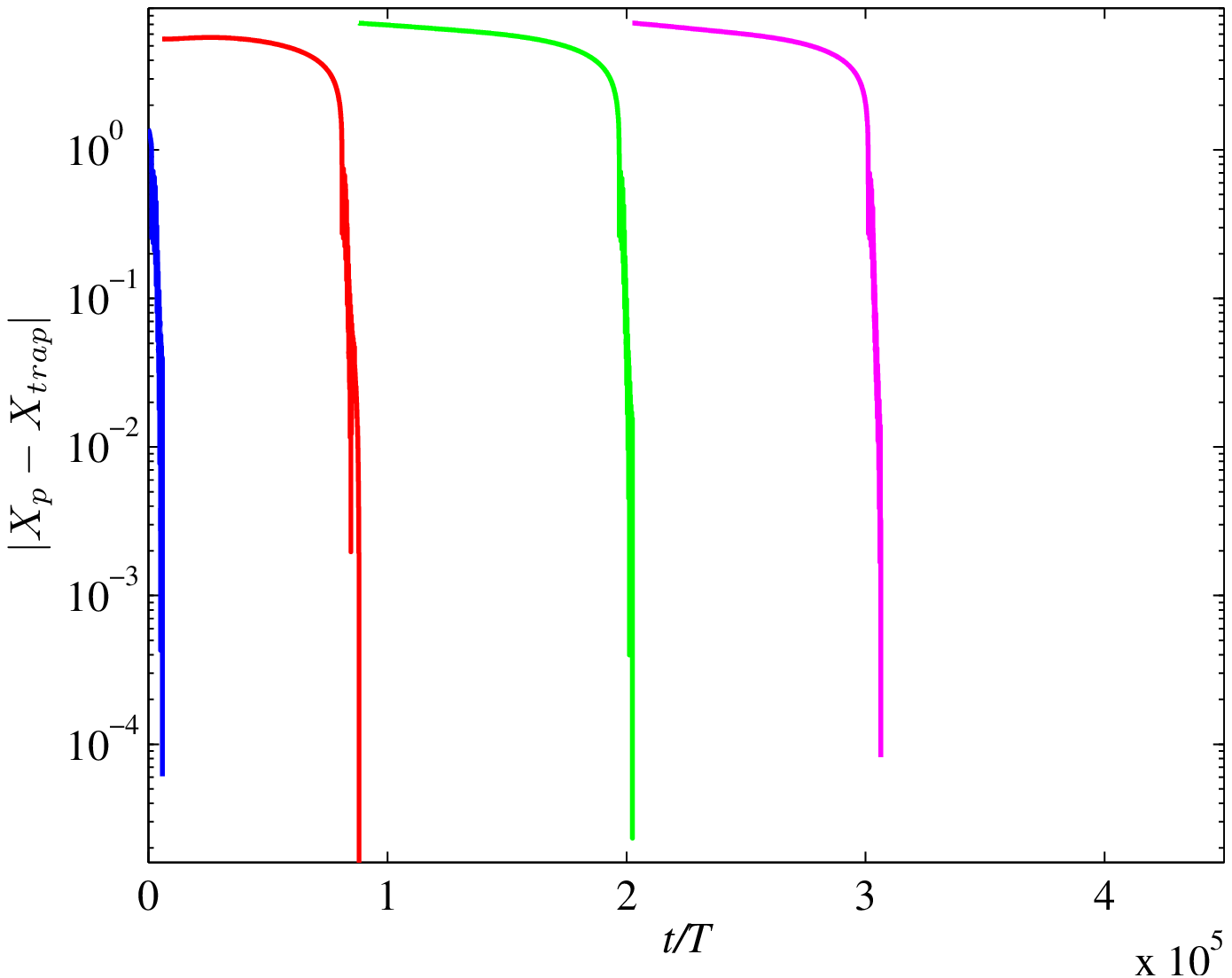}
   }\\
   \caption{(a) Inertial particle trajectory in a zigzag arrangement. (b) Distance between inertial particle($\xpart$) and trapping points
   ($\bbf{X}_{trap}$) of each respective oscillator.}\label{fig:stag}
\end{figure}

As suggested by these results, the overall transport speed can be improved if one can develop trajectories that avoid transport on a streamline near
the line of symmetry and in weak streaming zones created by obstacles. This can be ensured by arranging the oscillating cylinders in staggered
arrangement, as in figure \ref{fig:stag}(a). This strategy allows a particle to be transported approximately from $(-9R,0)$ to $(9R,0)$ with three
hundred thousand oscillations. Figure \ref{fig:stag}(b) indicates that the slow migration intervals have been significantly reduced by the staggered
arrangement.

\section{Conclusions} \label{sec:conclusion}
The transport of an inertial particle in a viscous streaming flow generated by multiple oscillating cylinders has been investigated by integrating
particle trajectories in a flowfield obtained by high-fidelity numerical simulation. With a controlled sequence of starting and stopping the
oscillation of cylinders, an inertial particle is transported from the streaming cell generated by one oscillating cylinder to the next in the
sequence. It has been shown that particles can be predictably transported and trapped in a wide variety of oscillator configurations. The staggered
arrangement minimizes the slow migration intervals.

Overall, we have shown that viscous streaming flows inside various arrangements of oscillating cylinders enable an effective mechanism for particle
manipulation. It should be noted that, for oscillation frequencies on the order of 100 kHz (at the upper range of frequencies studied by
\citet{hilgenfeldt:5j} for the oscillating bubble actuator), the transport times are on the order of three seconds. While this does not match the
rapid throughput of other particle sorting techniques, the strategy demonstrated here could be useful for more precise manipulation. In our ongoing
work, we are investigating the enhanced streaming topology that can be obtained with simultaneous oscillators to generate faster trajectories. We are
also exploring the potential to use this strategy for selective sorting of particles of various sizes and densities.


%

\end{document}